\documentstyle[psfig,12pt]{article}
\input psfig.sty
\def\reference{\parskip 0pt\par\noindent\hangindent 0.5 truecm}

\begin{document}
%
%
\title{Observations from Australasia using the Gravitational Microlensing Technique}
%


\author{Philip Yock$^{1}$ 
} 

\date{}
\maketitle

{\center
$^1$ Faculty of Science, University of Auckland, Auckland, New Zealand\\p.yock@auckland.ac.nz\\[3mm]
}

%
\begin{abstract}
The new astronomical technique of gravitational microlensing enables measurements of high precision to be made in certain circumstances. Useful advances have been made in the fields of galactic astronomy, stellar astronomy and planetary science. The technique is best suited to the southern sky, and several observations have been made from Australasia. A sample of these observations is described here. A case is also made for a telescope at the Antarctic dedicated to gravitational microlensing. 
\end{abstract}


\bigskip

%
%

\section{Introduction}
The new technique of gravitational microlensing has developed rapidly since its introduction to astronomy in 1993. During that year the MACHO group in Australia, and the EROS and OGLE groups in Chile, made the first observations (Alcock et al. 1993; Aubourg et al. 1993; Udalski et al. 1993). All three groups observed the apparent brightening of a distant star caused by the lens-like effect of a closer, collinear star, as had been predicted by Einstein (1936). The brightening was found to follow the characteristic achromatic light curve calculated by Paczynski (1986) as the stars moved into, and out of, collinearity. Paczynski had proposed using the technique to search for brown dwarfs in the dark halo of the Galaxy through their lensing effect on stars in the Magellanic Clouds. Subsequently, other uses for the technique were proposed, including studies of galactic structure, stellar spectra, stellar atmospheres and extra-solar planets. Other groups were formed to exploit these applications, some with telescopes in Australasia. 

In this paper, the main observations that have been carried out from Australasia to date are described. A proposal to include Antarctica as an observing site in the future is also described. For the most part the content follows an invited talk given by the author to the Joint Meeting of the Astronomical Society of Australia and the Royal Astronomical Society of New Zealand that was held at Sydney in July 1999. 

The observations from Australasia have been made by groups known as MACHO, GMAN, PLANET, MPS, and MOA. All the fields mentioned above have been studied, viz. dark matter, galactic structure, stellar spectra, stellar atmospheres and extra-solar planets. These are dealt with in turn in the paper. Most of the groups mentioned above maintain web sites on their work. These are at: \\
MACHO: \textit{http://wwwmacho.mcmaster.ca/} \\ 
PLANET: \textit{http://thales.astro.rug.nl/}$\sim$\textit{planet/} \\ 
MPS: \textit{http://bustard.phys.nd.edu/MPS/} \\
MOA: \textit{http://www.phys.vuw.ac.nz/dept/projects/moa/} \\  
EROS: \textit{http://www.lal.in2p3.fr/recherche/eros} \\ 
OGLE: \textit{http://www.astrouw.edu.pl/}$\sim$\textit{ftp/ogle/index.html}. \\
The last two groups, EROS and OGLE, are the Chilean based ones. The field of gravitational lensing as a whole, including its history dating back to 1804, was comprehensively reviewed recently by Wambsganss at  \textit{http://www.livingreviews.org/Articles/Volume1/1998-12wamb/index.html}.  
   
\section{Dark Matter}
The study of dark matter, and in particular brown dwarfs, was the first application of the gravitational microlensing technqiue to be proposed, and it motivated the early observations by the pioneering groups in the field, viz. MACHO, EROS and OGLE. The technique is particularly suited to the study of brown dwarfs or similar dark objects because it does not rely on optical detection.

Paczynski (1986) showed that, if the Galactic halo is composed of brown dwarfs, then approximately one star in a million in the Magellanic Clouds would be lensed at any one time by a brown dwarf for a duration $\sim140\sqrt{M/M_{\odot}}$ days. Here $M$ denotes the mass of the brown dwarf, which is $<0.08M_{\odot}$. Hence the expected duration for these lensing events is $\sim$40 days or less. The MACHO group has now monitored several million stars in the Magellanic Clouds for several years. Amongst their database of LMC stars to 1997, no events were found with duration less than 20 days. A similar result was obtained by the EROS group working from Chile. These results were combined to yield a $95\%$ confidence upper limit $\sim10\%$ on the halo-mass-fraction of low-mass brown dwarfs with masses in the range $(10^{-7}-10^{-2})M{_\odot}$ (Alcock et al. 1998). This result eliminates low-mass brown dwarfs, planets that have been ejected from planetary systems, and any other compact non-luminous objects in the stated mass range as a significant component of halo dark matter for standard models of the halo.

The MACHO and EROS groups extended their searches to include events of longer duration corresponding to brown dwarfs and other non-luminous objects with masses in the range $(0.01-1)M{_\odot}$. The EROS group did not detect any events caused by halo objects in this mass range in the direction of the SMC (Afonso et al. 1999a). The MACHO group has, however, recorded several events in this mass range in their database towards the LMC, corresponding to a halo fraction of $\sim50\%$ of objects with mass $\sim0.5M{_\odot}$ (Alcock et al. 1997a). The result, which is marginally consistent with the above result of the EROS group, was generally unexpected. It has proven difficult to interpret.

The above results are shown collectively in Figure 1. This figure also includes data from Gilmore and Unavane (1998) and the MOA group (Abe et al. 1999) that were obtained by surface photometry of external galaxies. The surface photometry data apply to main-sequence stars, i.e. to red dwarfs. Similarly restrictive limits on the abundance of red dwarfs in the Galactic halo have been obtained from studies of the Hubble deep field (Graff \& Freese 1996; Flynn, Gould \& Bahcall 1996).   

It is clear that the objects being detected by the MACHO group cannot be red dwarfs. Two interpretations of the data have been proposed. The first assumes a halo comprised mainly of old white dwarfs (Alcock et al. 1997a) or of primordial black holes (Nakamura 1998). The white dwarf hypothesis requires the initial mass function to be strongly peaked at $\sim2M{_\odot}$, to enable the white dwarfs to cool sufficiently to escape detection in other surveys (Chabrier et al. 1996), and it leaves unexplained the metal enrichment that would be expected to accompany them (Gibson and Mould 1997). The primordial black hole hypothesis requires their mass function to be peaked at $\sim0.5M{_\odot}$, surprisingly close to the mass of a normal star. Both these interpretations may be tested in the future by independent observations. Direct observations of white dwarfs should be possible if they are a major component of the halo (Chabrier 1999), and gravitational mergers of primordial black holes may be detectable if primordial black holes are a major component (Nakamura 1998).

The second type of interpretation that has been proposed for the MACHO events has potential implications for galactic structure. Sahu (1994) proposed that foreground stars in the LMC may lens background stars in the LMC, a process known as 'self lensing'. Initially, the expected event rate for self-lensing was expected to be too low to account for the observations (Gould 1995a). However, examples of self-lensing were subsequently detected (see section 5.2 below), and these lend weight to Sahu's hypothesis. Modified models of the LMC have been proposed that may accommodate a high self-lensing rate (Aubourg et al. 1999; Salati et al. 1999; Weinberg 1999), and methods have been proposed to test such models (Zhao 1999a, 1999b, 1999c). In a related proposal, Evans et al. (1998) proposed significant flaring and/or warping of the disk of the Milky Way to account for the MACHO events, a possibility that may also be tested by direct observation. 

In summary, the microlensing experiments have shown conclusively that Galactic dark matter is not comprised of brown dwarfs. Not all the data that have been obtained to date have been explained yet. Possibilities have been proposed, including old white dwarfs, primordial black holes, or modifications to current models of galactic structure. These may all be tested through independent observations.

\section{Galactic Bar}
The OGLE and MACHO groups included the Galactic bulge in their original targets for gravitational microlensing. Both groups reported a significantly higher event rate for microlensing than was expected (Udalski et al. 1994; Alcock et al. 1995). The high rate can be explained by assuming the Galactic bulge is bar shaped (Zhao, Rich \& Spergel 1996). In this model, stars from the near side of the bar lens stars from the far side. Supporting evidence for the model was obtained by the OGLE group who observed a systematic trend in the apparent magnitudes of red clump stars across the bulge (Stanek et al. 1997). These authors found that the  microlensing and red clump data can be jointly fitted by a bar inclined at $20^{\circ}-30^{\circ}$ to our line of sight. These results are consistent with earlier hints of a barred structure (e.g. Blitz \& Spergel 1991).

\section{Stellar Spectra}
A beautiful application of gravitational microlensing uses the amplification of the effect to effectively increase the size of a telescope to carry out observations of faint objects that would otherwise be impossible. The feasibility of this application has been ably demonstrated by Lennon et al. (1996, 1997) and by members of the MACHO group (Minniti et al. 1998). Minniti et al. measured the abundance of the rare element lithium in a main sequence bulge star using Keck I when the star was magnified by $\sim$1 magnitude, thereby effectively converting the diameter of the telescope to 15m. The microlensing event that was used for this application was MACHO-97-BLG-45, the forty-fifth event found towards the galactic bulge during 1997 by the MACHO group. 

This application of gravitational microlensing utilises the achromaticity of the effect, and also the predictability of the time of peak amplification once an event has commenced. The latter feature allows the requisite telescope scheduling to be readily carried out. The above lithium abundance measurement would not have been possible using conventional techniques and existing telescopes. Further such measurements are expected to determine the chemical composition and enrichment history of the bulge.      

\section{Stellar Atmospheres}
The study of stellar atmospheres by the gravitational microlensing provides an interesting application of the technique. Two methods are possible. Examples of each are given below.

\subsection{MACHO-1995-BLG-30}
During July 1995 the MACHO group reported a microlensing event in progress with unusual properties. The star being lensed was a red giant, and the lens trajectory was predicted to transit it. Photometric and spectroscopic observations of the event were requested from several locations, and these were carried out (Alcock et al 1997b).  The light curve for the event is shown in Figures 2 and 3. These include data from Australia, Chile, Israel and New Zealand. 

The above data for event MACHO-95-BLG-30 clearly show the effect of the finite size of the source star. As the lens transits the face of the source star, different parts of it are preferentially amplified. Limb darkening and stellar spots on the source star may then be detected (Heyrovsky, Sasselov \& Loeb 1999). For event MACHO-1995-BLG-30 an improved fit to the data was in fact obtained with a limb darkened model of the star. This is shown in Figure 3. Spectra taken during the event, shown in Figure 4, also showed some variation as the lens transited the source star. These are presently being analysed for the purpose of constructing a model atmosphere of the source star in this event (Heyrovsky \& Sasselov 1999).

\subsection{MACHO-98-SMC-1}
This event illustrates the second technique by which stellar atmospheres may be probed using gravitational microlensing. MACHO-98-SMC-1 was monitored by several groups including the PLANET group (Albrow et al. 1999). This group operates the network of telescopes shown in Figure 5. It enables almost continuous surveillance around the clock, weather permitting, of microlensing events. The lens for MACHO-98-SMC-1 was a binary star. Binary lenses can produce light curves that differ markedly from the single peaked structure shown in Figure 2 for a single lens. Figure 6 shows the light curve obtained by the PLANET group for MACHO-98-SMC-1. The asymptotic peaks occur when the source star crosses a 'caustic'. This is a linear region formed by the binary lens where the amplification is formally infinite (Alcock et al. 1999). The shape of the caustic determined by the PLANET group for MACHO-98-SMC-1 is shown in Figure 7.

The formally infinite amplification at a caustic is suppressed by the finite size of the source star. As the caustic sweeps across the face of the source star, different regions are preferentially amplified. This has two consequences. First, the crossing time can be measured and used to constrain the geometry of the event. Second, limb darkening in the source star can be detected. The first of these goals was achieved by the PLANET group for MACHO-98-SMC-1. They found a crossing time of $\sim$ 4.25 hr for the event, implying the lens was in the SMC (Albrow et al. 1999). The event was an example of the so-called 'self-lensing' process described in section 2, and the lens was not in the Galactic halo.

By combining the PLANET data for MACHO-98-SMC-1 with data from the EROS, MACHO, GMAN and MPS groups, a detection of limb darkening for the source star in five passbands ranging from I to V was obtained (Afonso et al. 1999b). Spectroscopy and photometry of the source star yielded an A6 dwarf classification. Thus, in this instance, limb darkening on a dwarf star some 60 kpc distant was detected! This highlights the extraordinary capability of the microlensing technique. Because the source star for this event was in the SMC, it is likely to be metal-poor. This represents the first measurement of limb darkening of an A type metal-poor star.                   

\section{Extra-Solar Planets}
The study of extra-solar planetary systems is one of the more intriguing applications of the gravitational microlensing technique. This application was first proposed by Mao and Paczynski (1991). Several refinements have since been discussed (Gould \& Loeb 1992; Bennett \& Rhie 1996; Gaudi \& Gould 1997; Wambsgnass 1997; Griest \& Safizadeh 1998; Gaudi, Naber \& Sackett 1998; Di Stefano \& Scalzo 1999). The technique may be likened to Rutherford's experiment on atomic structure as shown in Figure 8. Whereas Rutherford used $\alpha$ particles to probe atomic structure, gravitational microlensers use photons to probe planetary structure, by searching for deviations from the light curve produced by a single lens. The technique is promising because, as is indicated in the figure, a typical 'impact parameter' in a microlensing event is a few AU.  In contrast, the complementary Doppler technique enjoys greatest sensitivity for planets at orbital radii $\ll$1 AU (Marcy \& Butler 1998). Applications of the microlensing technique by the PLANET, MOA and MPS groups to events OGLE-98-BLG-14, MACHO-98-BLG-35 and MACHO-97-BLG-41 respectively are described below.

\subsection{OGLE-98-BLG-14}
The PLANET group made approximately 400 observations of OGLE-98-BLG-14 over a period of $\sim100$ days (Sackett 1999). The resulting light curve did not differ obviously from that of a single lens. A comparison was made between the observed light curve and the expected light curves for lenses with and without planets to place constraints on possible planets orbiting the lens star. The results are most easily expressed in terms of the Einstein radius $R_E$ for the event, and the planet-to-star mass ratio $\epsilon$.\footnote{The Einstein radius $R_E$ for bulge events is $\sim$ a few AU, although it is generally not known precisely on an event-by-event basis.} It was found that planets with $\epsilon>10^{-3}$ and orbital radii in the range (0.7-1.2)$R_E$ could be excluded with 100\% confidence, and that planets with $\epsilon>10^{-3}$ and orbital radii (0.6-1.5)$R_E$ could be excluded with 75\% confidence (Sackett 1999). These results strongly restrict the possible presence of Jupiter-like planets orbiting the lens star.  

\subsection{MACHO-98-BLG-35}
A significant refinement to the original proposal for planet hunting by Mao and Paczynski (1991) was made by Griest and Safizadeh (1998). They found that in microlensing events with high peak amplification, $\geq10$, a planet always  perturbs the light curve near its peak. The calculated perturbation was found to be detectable with essentially 100\% probability for Jupiter-like planets, and, depending on the geometry of the event, detectable with finite probability for lighter planets. Because the time of peak magnification of a microlensing event is generally known in advance, the finding by Griest and Safizadeh appeared to offer a systematic strategy for detecting planets.  

An opportunity arose to test the above strategy with event MACHO-98-BLG-35. This  reached a peak magnification $\sim80$. The peak of the event was monitored by the MPS and MOA groups (Rhie et al. 1999). Their light curves are shown in Figure 9. These include the best fit to the data assuming a lens with and without a planet. The parameters for the best fit with a planet were obtained with a planet mass ratio $\epsilon=7\times10^{-5}$ and an orbit radius of $1.35R_E$. Assuming a typical value for the lens mass $\sim0.3M_{\odot}$, these parameters correspond to a planet with mass between about that of Earth and about twice that of Neptune at an orbit radius of a few AU. The formal significance of the detection is at the $\sim4.5\sigma$ level.                

Figure 10 shows the same data as Figure 9 but plotted as a ratio to the best fit single lens light curve. Fig. 11 shows exclusion regions, at the $6.3\sigma$ confidence level, for planets with various masses. Jupiter-like and Saturn-like planets are excluded with quite high confidence. Peale (1997) has proposed the existence of planetary systems in which Jupiter and Saturn are replaced with Neptune-like planets. It is possible that this observation represents the first detection of such a system. 

The PLANET collaboration did not monitor MACHO-98-BLG-35 very frequently near its peak (see PLANET homepage - section 1). Such monitoring could have assisted greatly to check the above interpretation of the event. As a check on the quality of the MOA and MPS data we include here (Figures 12 and 13) the light curves by MOA of nearby stars of similar colour, crowdedness and magnitude as MACHO-98-BLG-35, and in Figure 14 the MPS data as a function of air-mass.

\subsection{MACHO-97-BLG-41}
This event was monitored by the MACHO and GMAN groups quite extensively. They found that single-lens and binary-lens models could not reproduce the data (Alcock et al. 1999). The event was also observed by the MPS group in association with a Wise Observatory team (Bennett et al. 1999). They have reported a model of the event in which the lens appears to be successfully treated as a planet orbiting a binary star. According to this model, the mass ratio of the binary star system is 3.8:1 and the stars are most likely to be a late K dwarf and an M dwarf with a separation of about 1.8 AU. A planet of about 3 Jupiter masses orbits this system at a distance of about 7 AU. 

The above result is the first detection of a giant planet by the gravitational microlensing technique. It suggests that such planets may be common in short period binary systems. Further data by other groups are available for the event that were not included in the above modeling. It will be interesting to see if a global analysis can pin down the parameters of the system more tightly. In any case, the above result signifies a major development in planetary science, whilst further highlighting the potential of the gravitational microlensing technique.

\section{Gravitational Microlensing from the Antarctic}
To realise the full potential of the gravitational microlensing technique it is necessary to monitor millions of stars with good photometric accuracy at a sampling rate of a few observations per hour in several passbands. The existing network of southern survey and follow-up telescopes (MACHO, OGLE, EROS, GMAN, PLANET, MPS and MOA) do a relatively good job of monitoring the Galactic bulge and the Magellanic Clouds. Further improvement may be expected to occur soon as new image subtraction techniques with better photometric accuracy are refined and incorporated (Alard \& Lupton 1998; Alard 1999). 

A quantum leap might be realised with a telescope at the Antarctic. The idea has been raised before  (Sahu 1998; Muraki et al. 1999). Such a telescope could monitor southern fields essentially continuously, thus avoiding the not inconsiderable difficulties associated with combining data from different groups using different telescopes and different passbands, and working under different seeing conditions. Losses of data due to inclement weather would also be less serious. To monitor the complete peak of a typical high magnification event from the Antarctic would require good weather for a few days in one location only. Presently, good weather is required simultaneously in Chile, in Australasia and in South Africa.\footnote{This is like asking for five days without rain for a cricket test in England.} By observing in the infrared, and taking advantage of the exceptionally dry conditions at the Antarctic, one could extend the present measurements to include the centre of the Galaxy. Gould has pointed out this would increase the total event rate (Gould 1995b). A 2-m class telescope would be the preferred option to extend the current observations being made with 1-m class instruments.

US and Australian groups have already made considerable progress towards the development of the Antarctic for infrared astronomy (Burton 1996). Observations have been made from the South Pole which confirm its excellent characteristics at infrared wavelengths, and site-testing is in progress at Dome-C (see Figure 15) which promises to be even superior (Burton 1999). In view of the above, Dome-C would seem to be a promising site for future development of gravitational microlensing.    

\section{Conclusions}
Gravitational microlensing is a useful technique in astronomy. Several new results have been obtained using the technique, including $ \\ 
\bullet\ elimination\ of\ brown\ dwarfs\ as\ halo\ dark\ matter \\
\bullet\ evidence\ for\ a\ bar\ in\ the\ Galaxy \\ 
\bullet\ spectroscopy\ of\ distant\ faint\ stars \\
\bullet\ measurements\ of\ limb\ darkening\ on\ distant\ stars \\
\bullet\ evidence\ for\ stars\ without\ Jupiter\!-\!like\ planets \\
\bullet\ evidence\ for\ a\ low\!-\!mass\ extrasolar\ planet \\
\bullet\ evidence\ for\ a\ planet\ orbiting\ a\ binary\ star\\$ 
Many of these results were enabled by data from the original survey groups, MACHO, OGLE and EROS, being made widely accessible. In the future, the Antarctic offers a promising site for extending the technique.

%
%






\section*{Acknowledgements}
The author thanks Ian Bond, Kem Cook and Nick Rattenbury for commenting on a draft version of the paper, Karen Pollard and Penny Sackett for supplying Figures 5-7, and Michael Burton for Figure 15.         

\section*{References}


\reference Abe, F. et al. 1999 AJ 118, 261
\reference Afonso, C. et al. 1999a A\&A 344, 63
\reference Afonso, C. et al. 1999b ApJ submitted, astro-ph/9907247 
\reference Alard, C. \& Lupton, R. 1998 ApJ 503, 325
\reference Alard, C. 1999 astro-ph/9903111
\reference Albrow, M. et al. 1999 ApJ 512, 672
\reference Alcock, C. et al. 1993 Nature 365, 621 
\reference Alcock, C. et al. 1995 ApJ 445, 133
\reference Alcock, C. et al. 1997a ApJ 486, 697
\reference Alcock, C. et al. 1997b ApJ 491, 436
\reference Alcock, C. et al. 1998 ApJ 499, L9
\reference Alcock, C. et al. 1999 ApJ submitted, astro-ph/9907369 
\reference Aubourg, E. et al. 1993 Nature 365, 623
\reference Aubourg, E. et al. 1999 A\&A submitted, astro-ph/9901372  
\reference Bennett, D. \& Rhie, S. 1996 ApJ 472, 660
\reference Bennett, D. et al. 1999 Nature submitted, astro-ph/9908038
\reference Blitz, L. \& Spergel, D. ApJ 379, 631 
\reference Burton, M. 1996 PASA 13, 2
\reference Burton, M. 1999 PASA submitted
\reference Chabrier, G., Segretain, L. \& Mera, D. 1996 ApJ 468, L21
\reference Chabrier, G. 1999 in Third Stromlo Symposium, ed. B. Gibson, T. Axelrod and M. Putman (ASP Conf. Series Vol. 165) 399 
\reference Di Stefano, R. \& Scalzo, R. 1999 ApJ 512, 579
\reference Evans, N. et al. 1998 ApJ 501, L45
\reference Einstein, A. 1936, Science 84 506
\reference Flynn, C., Gould, A. \& Bahcall, J. 1996 ApJ 466, L55 
\reference Gaudi, B., Naber, R. \& Sackett, P. 1998 ApJ 502, L33
\reference Gaudi, B. \& Gould, A. 1997 ApJ 486, 85
\reference Gibson, B. \& Mould, J. 1997 ApJ 482, 98
\reference Gilmore, G. \& Unavane, M. 1998 MNRAS 301, 813
\reference Gould, A. 1995a ApJ 441, 77
\reference Gould, A. 1995b ApJ 446, L71
\reference Gould, A. \& Loeb, A. 1992 ApJ 396, 104
\reference Graff, D. \& Freese, K. 1996 ApJ 456, L49
\reference Griest, K. \& Safizadeh, N. 1998 ApJ 500, 37
\reference Heyrovsky, D., Sasselov, D. \& Loeb, A. 1999 ApJ submitted, astro-ph 9902273 
\reference Heyrovsky, D \& Sasselov, D. 1999 ApJ submitted, astro-ph/9906024
\reference Lennon, D., Mao, S., Fuhrmann, K. \& Thomas, G. 1996 ApJ 471, L23
\reference Lennon, D. et al. 1997 the Messenger 90, 30
\reference Mao, S. \& Paczynski, B. 1991 ApJ 374, L37
\reference Marcy, G. \& Butler, R. 1998 ARA\&A 36, 57
\reference Minniti, D. et al. 1998 ApJ 499, L175
\reference Muraki, Y. et al. 1999 Suppl. Prog. Theor. Phys. 133, 233 
\reference Nakamura, T. 1998 Physics Reports 307, 181
\reference Paczynski, B. 1986 ApJ 304, 1 
\reference Peale, S. 1997 Icarus 127, 269
\reference Rhie, S. et al. 1999 ApJ submitted, astro-ph/9905151
\reference Sackett, P. 1999 in ESO Astrophysics Symposia (Springer Verlag) in press
\reference Sahu, K. 1994 Nature 370, 275
\reference Sahu, K. 1998 in Astrophysics from Antarctica, ed. G. Novak and R. Landsberg (ASP Conf. Series) 179 
\reference Salati, P. et al. 1999 A\&A submitted, astro-ph/9904400
\reference Schechter, P., Mateo, M. \& Saha, A. 1993 PASP 105, 1342 
\reference Stanek, K. et al. 1997 ApJ 477, 163
\reference Udalski, A. et al. 1993 Acta Astron 43, 289
\reference Udalski, A. et al. 1994 Acta Astron. 44, 165
\reference Wambsganss, J. 1997 MNRAS 284, 172  
\reference Weinberg, M. 1999 ApJ submitted, astro-ph/9905305
\reference Zhou, H., Rich, R. \& Spergel, D. 1996 MNRAS 282, 175 
\reference Zhou, H. 1999a ApJ in press, astro-ph/9906126
\reference Zhou, H. 1999b ApJ in press, astro-ph/9906214
\reference Zhou, H. 1999c ApJ submitted, astro-ph/9907191

\section*{Figure Captions}
1. Upper limits to mass fractions of halo objects of various masses determined by Alcock et al. (1998a) - thin line, by Afonso et al. (1999a) - thick line, by Gilmore \& Unavane (1998) - thin dashed line, and by Abe et al. (1999b) - thick dashed line. The possible detection is by Alcock et al. (1997a).  The limits by Gilmore \& Unavane, and by Abe et al., were obtained by surface photometry of external galaxies. They apply to main-sequence stars only. \\ \\
2. Light curve of gravitational microlensing event MACHO-1995-BLG-30. The additional schematic relates the scale of the lens's Einstein radius to the angular size of the source star, and indicates transit of the lens across the source face. The Einstein radius $R_E$ is the impact parameter of the light at the lens plane assuming the lens to be perfectly aligned with the source. \\ \\
3. Peak structure of microlensing event MACHO-1995-BLG-30, showing the best standard microlensing fit to the data (dashed curve), and an extended source microlensing fit incorporating source limb-darkening (solid curve).\\ \\
4. Spectral observations of event MACHO-1995-BLG-30 plotted on the source plane. The relative sizes of the star (solid circle) and the lens's Einstein radius (dotted circle) are plotted to scale, in units of $R_E$. The solid points show the position of the lens at the times when spectra were taken. The observatories where the spectra were taken are indicated by tick marks.\\ \\
5. Network of telescopes operated by the PLANET group.\\ \\
6. Light curve by the PLANET group for event MACHO-98-SMC-1. The data are from SAAO 1-m (circles), the CTIO 90-cm (squares), the CTIO-Yale 1-m (triangles), and the Canopus 1-m (asterisks) telescopes. The inset covers 0.6 days, corresponding to less than one tick mark on the main figure. The data are binned by day on the main figure. Two fits to the data are shown. These are discussed in Albrow et al. (1999). As is apparent from the figure, the models have similar crossing times.\\ \\
7. Caustic crossing geometry of event MACHO-98-SMC-1 as determined by Albrow et al. (1999). The diamond-shaped curve is the caustic. The thick solid line and the two thinner parallel lines indicate the trajectory and size of the source star. The two components of the binary are shown by circles whose relative sizes are proportional to their masses. The tick marks are in units of the Einstein crossing time $\hat{t}$ = 108.4 days.\\ \\
8. Comparison of Rutherford scattering and gravitational microlensing. The longitudinal scale has been compressed in the lower illustration.\\ \\
9. Peak structure of event MACHO-98-BLG-35 as determined by the MPS and MOA groups (Rhie et al. 1999). The pale curve is the best fit to the data for a lens without a planet, and the heavy curve is the best fit for a lens with a single planet with mass fraction $\epsilon=7\times10^{-5}$ and orbit radius = 1.35 Einstein radii.\\ \\
10. The same data as in Figure 9 but plotted as a ratio to the best fit to the data for a lens without a planet.\\ \\  
11. Excluded regions in the lens plane in units of $R_E$ for planets of various mass fractions $\epsilon$. The horizontal axis denotes the track of the source star, from right to left.\\ \\
12. Light curves of neighbouring stars to MACHO-98-BLG-35 of similar magnitudes, colours and crowdedness. These data were obtained by the MOA group on the first night shown in Figures 9 and 10. The error bars are those returned by the DoPHOT point-spread-function fitting routine (Schechter, Mateo \& Saha 1993). The error bars shown in Figures 9 and 10 have had an additional $1\%$ added to them in quadrature, to account for possible flat-fielding and normalisation errors that are not allowed for in the DoPHOT routine. The enlarged error bars were used in the analysis of the event described above and in Rhie et al. (1999).\\ \\
13.  Similar to Figure 12 above, but for the second night shown in Figures 9 and 10.\\ \\
14. The magnification ratio from the best fit single lens light curve for MACHO-98-BLG-35 is plotted as a function of airmass for the MPS data taken within the week centered on the time of maximum magnification.\\ \\
15. Map of Antarctica from Burton (1996), showing the Plateau bases at the South Pole (USA) and Vostok (Russia) in relation to Australia and New Zealand. The high Plateau site being considered for a future Antarctic observatory lies inside the 3000-m elevation contour at Dome C. The South Pole is supplied through the US coastal station at McMurdo. The Australian coastal stations at Mawson, Davis and Casey, and the New Zealand Scott Base, are also shown.\\ \\

\end{document}